\newcommand{\bee}{\begin{equation}}
\newcommand{\ene}{\end{equation}}
\newcommand{\beea}{\begin{eqnarray}}
\newcommand{\enea}{\end{eqnarray}}

\documentclass[aps,preprint,pre,showpacs,superscriptaddress]{revtex4}
\usepackage{amsmath} %%%% To be able to use 'align' and 'subfigures', etc.
\usepackage{graphicx}% Include figure files
\usepackage{dcolumn}% Align table columns on decimal point
\usepackage{bm}% bold math
\begin{document}
\title{Characteristics of Plasmon Transmittivity Over Potential Barriers}
\author{M. Akbari-Moghanjoughi}
\affiliation{Faculty of Sciences, Department of Physics, Azarbaijan Shahid Madani University, 51745-406 Tabriz, Iran}

\begin{abstract}
In this paper we use the Schr\"{o}dinger-Poisson model to obtain a linear coupled pseudoforce system from which the wave function and the electrostatic potential of the free electron gas plasmon is deduced. It is shown that unlike for single particle wave function the plasmon wave function and corresponding electrostatic potential are characterized with two different wave numbers associated with two distinct characteristic length scales, namely, that of the single electron oscillation and of the collective Langmuir excitations. Interaction of plasmon with a rectangular potential step/well indicates features common with that of the ordinary single quantum particle. However, the two-tone oscillation character of the wave function and potential appear on the transmitted amplitude over the potential barrier/well. The plasmon propagation is found to have a distinct energy gap corresponding to the plasmon energy value of $\epsilon_g=\mu_0+2\epsilon_p$ below which no plasmon excitations occur. For instance, the zero-point plasmon excitation energy for Aluminium, is around $\epsilon_0\simeq 41.7$eV at room temperature, with the Fermi energy of $\epsilon_F\simeq 11.7$eV and plasmon energy of $\epsilon_p\simeq 15$eV. It is seen that for plasmon energies very close to the energy gap, i.e. where the two characteristic scales match ($k_1\simeq k_2$), the quantum beating effect takes place. The study of plasmon tunneling through the potential barrier indicates that the transmittivity has oscillatory behavior similar to that of a quantum particle tunneling through the potentials, but, with a characteristic two-tone oscillatory profiles. Current development can have a broad range of applications in plasmon transport through diverse free electron environments with arbitrary degeneracy and electron temperature. It also makes progress in rapidly growing nanotechnology and plasmonic fields.
\end{abstract}

\date{\today}

\maketitle

\section{Introduction}

\textbf{Plasmons are the simplest collective excitations of free electron systems such as plasmas, metals and nano-metallic structures \cite{rae,dib}. Technological advancement in the rapidly growing fields of optoelectronics \cite{haug,wooten}, integrated circuits \cite{hu}, plasmonic \cite{man1}, nanometallic design, quantum computing and other applications are all dependent on clear understanding of the collective electron-hole dynamics and mobilities in electronic devices  \cite{sarma,stern2,hwang,gupta,datta,markovich}. The quantum tunnelling and transmittivity over potential barriers is fundamental application of quantum mechanical phenomenon in modern devices such as tunnel diode \cite{seeg}, scanning tunneling microscope \cite{taylor,knight}, field effect transistors \cite{ion} and high speed semiconductor tunneling switches \cite{gi}. Quantum tunneling is also an essential ingredient in several astrophysical processes such as nuclear fusion, radioactivity and astrochemistry in interstellar clouds \cite{tri}. Moreover, the quantum transport phenomenon in metals uses the tunnelling concept to fully describe the nature of electron collisions and conductivity through the periodic lattice potential \cite{kit}. Quantum tunnelling is also the dominant electron-hole transport phenomenon in the very large scale (VLS) integrated circuit devices. The most simple tunnel configuration is made of micron sized gap between two metals called the Josephson junction which finds numerous applications in precision measurements and multijunction solar cells \cite{ash}. The quantum tunneling effect which a wave aspect of electron transport can take place in physical barriers and insulator gaps with typical sizes of $1$-$3$nm. In a tunnel diode however the barrier is the depletion layer in an n-p semiconductor juction. Therefore, by applying a forward bias the electrons can tunnel through the barrier making a significant current in the diode \cite{krane}.}

The field quantum plasmas has attracted much attention in recent years due to applications in quantum transport and plasmon excitation. The pioneering original contributions to this field include those of Fermi \cite{fermi}, Madelung \cite{madelung}, Hoyle and Fowler \cite{hoyle}, Chandrasekhar \cite{chandra}, Bohm \cite{bohm} Pines \cite{pines}, Levine \cite{levine}, Klimontovich and Silin \cite{klimontovich} and many others. These contributions and many recent developments on quantum plasmas \cite{man2,shuk1,haas1,brod1,mark1,man3,asenjo,brod2,mark2,ydj1,ydj2,dub1,shuk2,shuk3,shuk4,mark3,akbrevisit} not only have expanded our knowledge of quantum systems and their statistical behavior, but also have led to discovery of many interesting collective properties of dense and degenerate ionized environments unobserved in classical domain. The characteristic metallic properties of solids like the ground state energy of free electron system are well described by Fermi-Thomas model which rely on the quantum statistical features of fermionic systems. These properties are of fundamental value particularly to the field of solid state in describing the most basic collective behavior of metals such as specific heat and electric or heat transport \cite{kit,ash}. For instance, the most fundamental electron transport model in solids is due to Sommerfeld which uses the quantum statistic effects in order to takes into account the collective plasma oscillations. However, even the most advanced theories of the quantum transport in metallic compounds present today, approach the many electron effects by taking the single-electron $k$-state in the plane-wave propagations. While successful to some extent, they are not successful in bringing both the single and many-electron oscillation character of the fermion system in a pleasant unified picture. This is due to the fact that the dual $k$-state picture can not be probed by the conventional Schr\"{o}dinger equation which has been designed to cope with single-particle wave function. It is however well-known that full understanding of the physical properties of many electron system requires self-consistent solution of both Schr\"{o}dinger and Poisson equations \cite{manfredi}. But, due to the complex mathematical nature of Schr\"{o}dinger-Poisson model only numerical methods are available in order to evaluate this system \cite{tan,tom,mui,sni,lux}. In this paper we present a simple method to reduce the linearized system to a coupled pseudoforce differential equations \cite{akbpseudo1,akbpseudo2} solution of which presents the plane wave structure of the full many electron system which reveals the dual length-scale character of the system in a single picture. Then by using this plane wave solution for the free plasmon excitation, similar to single particle case, we investigate  the interaction of a free electron gas with some simple external potential step and barrier and compare our results to tunneling effect through the potential by a single particle. The presentation is as follows. The basic physical model is presented in Sec. II. The generalized plane-wave solution is derived in Sec. III. Plasmon interaction with step potential is investigated in Sec. IV. Interaction with a rectangular potentia is given in Sec. V and conclusions are presented in Sec. VI.

\section{Physical Model}

We consider a one dimensional (1D) free electron gas with an arbitrary degree of nonrelativistic degeneracy, the chemical potential, $\mu$, and the equilibrium temperature, $T$. The evolution of such system is best described by the-Poisson model as \cite{akbnew}
\begin{subequations}\label{sp}
\begin{align}
&i\hbar \frac{{\partial \cal N }}{{\partial t}} =  - \frac{{{\hbar ^2}}}{{2m}}\frac{{\partial {{\cal N} ^2}}}{{\partial {x^2}}} - (e\phi-eV_0){\cal N}  + \mu(n,T){\cal N},\\
&\frac{{\partial {\phi ^2}}}{{\partial {x^2}}} = 4\pi e n,
\end{align}
\end{subequations}
in which ${\cal N} =\sqrt{n(x,t)}\exp[iS(x,t)]$ is the electron gas wave function with ${\cal N}{\cal N^*}=n(x,t)$ being the number density and $u(x,t)=(1/m)\partial S(x,t)/\partial x$ the electron fluid speed. Also, $V_0$ is any constant external potential, $\phi$ the electrostatic potential, $\hbar$ the reduced Planck constant and $m$ the electron mass. In the linear perturbation limit, when $\mu\equiv\mu_0$, however, the system (\ref{sp}) reduces to the following coupled linear pseudoforce system with the particular plasmon plane-wave solution of type ${\cal N}(x,t)=\psi(x)\exp(i\epsilon t/\hbar)$ with energy eigenvalue, $\epsilon$
\begin{subequations}\label{pf}
\begin{align}
&\frac{{d^2{\Psi(x)}}}{{d{x^2}}} + \Phi(x) = - 2 E \Psi(x),\\
&\frac{{d^2{\Phi(x)}}}{{d{x^2}}} - {\Psi}(x) = 0,
\end{align}
\end{subequations}
where we have used the normalization scheme $\Psi=\psi/\sqrt{n_0}$ with $n_0$ being the equilibrium number density of electron gas, $\Phi=e\phi$, $E=(\epsilon-\mu_0- eV_0)/2\epsilon_p$ ($\epsilon_p=\hbar\omega_p$ with $\omega_p=\sqrt{4\pi e^2 n_0/m}$ is the plasmon energy quanta) and $x=x/\lambda_p$ where $\lambda_p=2\pi/k_p$ with $k_p=\sqrt{2m\epsilon_p}/\hbar$ is the characteristic plasmon wavelength. The number density and pressure in (\ref{eos}) can also be rewritten in terms of familiar analytic polylog functions as follows \cite{ae}
\begin{equation}\label{pol}
{n(\mu_0,T)} =  - {N}{\rm{L}}{{\rm{i}}_{3/2}}\left[ { - {\rm{exp}}\left( {-\beta {\mu_0}} \right)} \right],\hspace{3mm}{P(\mu_0,T)} =  - \frac{N}{\beta}{\rm{L}}{{\rm{i}}_{5/2}}\left[ { - {\rm{exp}}\left( {-\beta {\mu_0}} \right)} \right],
\end{equation}
where $\beta=1/k_B T$ with $k_B$ being the Boltzmann constant and $N$ is defined as below
\begin{equation}\label{N}
{N} = \frac{2}{{\Lambda^3}} = 2\left( {\frac{{{m k_B T}}}{{{2\pi\hbar^2}}}} \right)^{3/2},
\end{equation}
with $\Lambda$ being the electron thermal de Broglie wavelength. Therefore, the equation of state of an isothermal free electron gas with arbitrary degeneracy may be written as \cite{ae}
\begin{equation}\label{eos}
P(\mu_0,T)=\frac{n(\mu_0,T)}{\beta}\frac{{\rm{L}}{{\rm{i}}_{5/2}}[ - \exp ({-\beta\mu_0})]}{{\rm{L}}{{\rm{i}}_{3/2}}[ - \exp ({-\beta\mu_0})]}.
\end{equation}

\section{Generalized Plane Wave Solution}

The general complex solution to the coupled linear system (\ref{pf}) is given below
\begin{subequations}\label{gs}
\begin{align}
&\Psi (x) = \frac{{\left( {k_2^2{\Psi _0} + {\Phi _0}} \right)\exp \left( {{\rm{i}}{k_2}x} \right) - \left( {k_1^2{\Psi _0} + {\Phi _0}} \right)\exp \left( {{\rm{i}}{k_1}x} \right)}}{{2\alpha }},\\
&\Phi (x) = \frac{{\left( {k_2^2{\Phi _0} + {\Psi _0}} \right)\exp \left( {{\rm{i}}{k_1}x} \right) - \left( {k_1^2{\Phi _0} + {\Psi _0}} \right)\exp \left( {{\rm{i}}{k_2}x} \right)}}{{2\alpha }},
\end{align}
\end{subequations}
where $\Psi_0$ and $\Phi_0$ are defined by the initial condition of plasmon excitations. The solutions (\ref{gs}) may be considered as the most general plane wave solution to plasmon excitations with two distinct characteristic wave numbers (normalized to $k_p$) as defined below
\begin{equation}\label{eks}
{k_1} = \sqrt {E_0 - \alpha },\hspace{3mm}{k_2} = \sqrt {E_0 + \alpha },\hspace{3mm}\alpha  = \sqrt {{E_0^2} - 1}.
\end{equation}
It is noted that the dual scale-length character of the free electron gas comes from the single-particle and collective interaction of the electrons \cite{akbpseudo2}. For a free electron wave we have $E_0=(\epsilon-\mu_0)/2\epsilon_p$ where $\epsilon$ is the plasmon energy and $\mu_0$ is the unperturbed chemical potential of the system. Note that there is a plasmon energy gap for $E_0<1$ or $\epsilon<\mu_0+2\epsilon_p$. The plasmon excitation gap energy is then $\epsilon_g=\mu_0+2\epsilon_p$. The typical value of room-temperature Fermi and plasmon energies for Aluminium are $\mu_0=\epsilon_F\simeq 11.7$eV and $\epsilon_p\simeq 15$eV, respectively. It can also be easily confirmed that the energy of free electron excitations given by (\ref{gs}) is $\epsilon=\mu_0+\epsilon_p(k_1^2+k_2^2)$. The quantized energy levels of collective electron gas excitations in an infinite potential well are obtained in Ref. \cite{akbnew}. Application of the constant external potential $V_0$ to the electron gas results in the same plane wave solution (\ref{gs}) with new energy eigenvalue $E=(\epsilon-\mu_0-eV_0)/2\epsilon_p=E_0-U$ with $U=eV_0/2\epsilon_p$ and $E_0$ denoting the normalized potential-free plasmon energy. The normalized energy $E_0\simeq 1$ or equivalently $\epsilon\simeq \mu_0+2\epsilon_p$ leads to the quantum beating effect in which $k_1\simeq k_2$ holds.

\section{Interaction with Potential Steps}

Using the generalized plane wave solution for linear plasmon excitation we now consider the interaction of these entities with a simple potential step of height/depth, $U$. The problem dealt with in a quite similar manner as with the particle penetrating into the (negative/possitive) potential region. The wave function is partly reflected and transmitted through the divided region. We consider the boundary to be at the origin $x=0$. The boundary conditions at $x=0$ which is the continuity of solutions and their derivatives at the boundary of the two regions, properly provide the values of $\Psi_0$ and $\Phi_0$ in (\ref{gs}), as follows
\begin{subequations}\label{bc1}
\begin{align}
&{\Phi _1}(0) = {\Phi _2}(0),\hspace{3mm}{\left. {\frac{{d{\Phi _1}(x)}}{{dx}}} \right|_{x = 0}} = {\left. {\frac{{d{\Phi _2}(x)}}{{dx}}} \right|_{x = 0}},\\
&{\Psi _1}(0) = {\Psi _2}(0),\hspace{3mm}{\left. {\frac{{d{\Psi _1}(x)}}{{dx}}} \right|_{x = 0}} = {\left. {\frac{{d{\Psi _2}(x)}}{{dx}}} \right|_{x = 0}}.
\end{align}
\end{subequations}
The wavefunction and electrostatic potential at the two regions can be written as follows
\begin{subequations}\label{wp1}
\begin{align}
&{\Phi _1}(x) = \frac{{\left( {{k_{21}^2}{\Phi _0} + {\Psi _0}} \right)\exp \left( {{\rm{i}}{k_{11}}x} \right) - \left( {{k_{11}^2}{\Phi _0} + {\Psi _0}} \right)\exp \left( {{\rm{i}}{k_{21}}x} \right)}}{{2{\alpha _1}}},\\
&{\Phi _2}(x) = \frac{{\left( {{k_{22}^2}{\Phi _0} + {\Psi _0}} \right)\exp \left( {{\rm{i}}{k_{12}}x} \right) - \left( {{k_{12}^2}{\Phi _0} + {\Psi _0}} \right)\exp \left( {{\rm{i}}{k_{22}}x} \right)}}{{2{\alpha _2}}},\\
&{\Psi _1}(x) = \frac{{\left( {{k_{21}^2}{\Psi _0} + {\Phi _0}} \right)\exp \left( {{\rm{i}}{k_{21}}x} \right) - \left( {{k_{11}^2}{\Psi _0} + {\Phi _0}} \right)\exp \left( {{\rm{i}}{k_{11}}x} \right)}}{{2{\alpha _1}}} \\
&+ r\frac{{\left( {{k_{21}^2}{\Psi _0} + {\Phi _0}} \right)\exp \left( { - {\rm{i}}{k_{21}}x} \right) - \left( {{k_{11}^2}{\Psi _0} + {\Phi _0}} \right)\exp \left( { - {\rm{i}}{k_{11}}x} \right)}}{{2{\alpha _1}}},\\
&{\Psi _2}(x) = t\frac{{\left( {{k_{22}^2}{\Psi _0} + {\Phi _0}} \right)\exp \left( { - {\rm{i}}{k_{22}}x} \right) - \left( {{k_{12}^2}{\Psi _0} + {\Phi _0}} \right)\exp \left( { - {\rm{i}}{k_{12}}x} \right)}}{{2{\alpha _2}}},
\end{align}
\end{subequations}
where $r$ and $t$ are the reflection and transmission amplitude, respectively and with the characteristic wavenumber and energy values given as
\begin{subequations}\label{un}
\begin{align}
&{\alpha _1} = \sqrt {{E_1^2} - 1},\hspace{3mm}{k_{11}} = \sqrt {{E_1} - {\alpha _1}},\hspace{3mm}{k_{21}} = \sqrt {{E_1} + {\alpha _1}},\\
&{\alpha _2} = \sqrt {{E_2^2} - 1},\hspace{3mm}{k_{12}} = \sqrt {{E_2} - {\alpha _2}},\hspace{3mm}{k_{22}} = \sqrt {{E_2} + {\alpha _2}},\hspace{3mm}{E_2} = {E_1} - {U}.
\end{align}
\end{subequations}
It is remarkable to find that, for any given values of $E\neq 1$, $k_{11}$ and $k_{12}$ are respective reciprocals of $k_{21}$ and $k_{22}$, that is, $k_{11}k_{21}=k_{12}k_{22}=1$. This leads to the simple normalized energy dispersion of $E=(1+k^4)/2k^2$ for the free electron plasmon excitations. Application of the continuity of the electrostatic potential function $\Phi(x)$ at the boundary confirms that, regardless of the initial values $\Phi_0$ and $\Psi_0$, the boundary conditions are always satisfied. Moreover, The continuity of the wavefunction $\Psi(x)$ at the boundary leads to $t=1+r$ a result which is also independent from the values of $\Phi_0$ and $\Psi_0$. The later result is comparable to that of quantum particle tunneling through the step potential. Continuity of potential derivative at the boundary, however, leads to the condition below
\begin{equation}\label{b1}
{\Phi _0} = \frac{{\left[ {{\beta}\left( {{k_{12}} - {k_{22}}} \right) - \left( {{k_{11}} - {k_{21}}} \right)} \right]{\Psi _0}}}{{{\beta}{k_{12}}{k_{22}}\left( {{k_{12}} - {k_{22}}} \right) - {k_{11}}{k_{21}}\left( {{k_{11}} - {k_{21}}} \right)}},
\end{equation}
a close inspection of which reveals that $\Psi_0=\Phi_0$ must hold, due to the following identity
\begin{equation}\label{id}
\beta  = \frac{{{\alpha _1}}}{{{\alpha _2}}} = \frac{{\left( {{k_{11}} - {k_{21}}} \right)\left( {1 - {k_{11}}{k_{21}}} \right)}}{{\left( {{k_{12}} - {k_{22}}} \right)\left( {1 - {k_{12}}{k_{22}}} \right)}}.
\end{equation}
Using the conditions above one may rewrite the wave function and the corresponding potentials in the following simplified forms
\begin{subequations}\label{wp1s}
\begin{align}
&{\Phi _1}(x) = \frac{{\left( {1 + k_{21}^2} \right){{\rm{e}}^{{\rm{i}}{k_{11}}x}} - \left( {1 + k_{11}^2} \right){{\rm{e}}^{{\rm{i}}{k_{21}}x}}}}{{2{\alpha _1}}},\\
&{\Phi _2}(x) = \frac{{\left( {1 + k_{22}^2} \right){{\rm{e}}^{{\rm{i}}{k_{12}}x}} - \left( {1 + k_{12}^2} \right){{\rm{e}}^{{\rm{i}}{k_{22}}x}}}}{{2{\alpha _2}}},\\
&{\Psi _1}(x) = \frac{{\left( {1 + k_{21}^2} \right){{\rm{e}}^{{\rm{i}}{k_{21}}x}} - \left( {1 + k_{11}^2} \right){{\rm{e}}^{{\rm{i}}{k_{11}}x}}}}{{2{\alpha _1}}} + r\frac{{\left( {1 + k_{21}^2} \right){{\rm{e}}^{ - {\rm{i}}{k_{21}}x}} - \left( {1 + k_{11}^2} \right){{\rm{e}}^{ - {\rm{i}}{k_{11}}x}}}}{{2{\alpha _1}}},\\
&{\Psi _2}(x) = t\frac{{\left( {1 + {k_{22}}^2} \right){{\rm{e}}^{{\rm{i}}{k_{22}}x}} - \left( {1 + {k_{12}}^2} \right){{\rm{e}}^{{\rm{i}}{k_{12}}x}}}}{{2{\alpha _2}}}.
\end{align}
\end{subequations}
On the other hand, the continuity of wave function derivative at the boundary leads to
\begin{equation}\label{wd1}
t = \frac{{\left( {1 - r} \right)\left[ {{k_{11}}(1 + {k_{11}^2}) - {k_{21}}\left( {1 + {k_{21}^2}} \right)} \right]}}{{\beta \left[ {{k_{12}}(1 + {k_{12}^2}) - {k_{22}}\left( {1 + {k_{22}^2}} \right)} \right]}}.
\end{equation}
Combination of (\ref{wd1}) with $t=1+r$ results in the following relations for reflection and transmission amplitudes
\begin{subequations}\label{tr}
\begin{align}
&r = \frac{{\left[ {{k_{11}}(1 + {k_{11}^2}) - {k_{21}}\left( {1 + {k_{21}^2}} \right)} \right] + \beta \left[ {{k_{22}}(1 + {k_{22}^2}) - {k_{12}}\left( {1 + {k_{12}^2}} \right)} \right]}}{{\left[ {{k_{11}}(1 + {k_{11}^2}) - {k_{21}}\left( {1 + {k_{21}^2}} \right)} \right] + \beta \left[ {{k_{12}}(1 + {k_{12}^2}) - {k_{22}}\left( {1 + {k_{22}^2}} \right)} \right]}},\\
&t = \frac{{2\left[ {{k_{11}}(1 + {k_{11}^2}) - {k_{21}}\left( {1 + {k_{21}^2}} \right)} \right]}}{{\left[ {{k_{11}}(1 + {k_{11}^2}) - {k_{21}}\left( {1 + {k_{21}^2}} \right)} \right] + \beta \left[ {{k_{12}}(1 + {k_{12}^2}) - {k_{22}}\left( {1 + {k_{22}^2}} \right)} \right]}}.
\end{align}
\end{subequations}
The reflectivity $R$ and transmittivity $T$ follow the relations $R=|r|^2$ and $T=1-R$.

Figure 1(a) depicts the schematic of the potential step of height $U=5$ as normalized to $2\epsilon_p$ to the right $x>0$. A plasmon with a normalized energy $E$ is assumed to be incident on the potential step from the left and after interaction partially transmit/reflected from the potential to the right/left regions. There is however an energy gap for propagation of the plasmon excitations in the normalized energy values of $E<1$ which corresponds to the zero-point plasmon excitation energy of $\epsilon_0=\mu_0+2\epsilon_p$ in a degenerate electron gas with chemical potential $\mu_0$. The zero-point energy depends on the electron gas number density and its temperature with a typical value of $\epsilon_0\simeq 41.7$eV for the Aluminium at room temperature. \textbf{A realistic potential step $U$ may be due to voltage bias applied across the electron gas, a step-like change of the chemical potential $\mu_0$ (Fermi energy for completely degenerate gas) in the border of epitaxially grown plasmonic crystals or a very narrow depletion region in a heavily doped p-n junction. However, treatment of more general quantum potentials, by using the scattering matrix method, is postponed for a future investigation. In Fig. 1(b) we have shown the negative potential step of depth $U=-5$ to the right $x>0$. Note that in the plasmon transmission over the potential steps of Figs. 1(a) and 1(b) there are exists a transmission energy gap corresponding to the ranges of $U-1<E<U+1$. It is also found that for the energy range $1<E<U-1$ the plasmon is damped in the positive potential step region. In practice an appropriately doped p-n junction may operate as an ultrafast electronic switch using a voltage bias as the potential step height tuning parameter.}

The reflectivity $R(E)$ and transmittivity $T(E)$ are shown for different values of the applied potential energy $U=eV_0/2\epsilon_p$ in two cases of positive and negative potential steps in Figs. 1(c) and 1(d), respectively. Figure 1(c) shows the reflection and transmission coefficients for the potential steps with different heights. The  reflectivity $R(E)$ and transmittivity $T(E)$ profiles are quite similar to that of a single particle quantum tunneling through the potential steps with same height, except that, they are shifted one energy unit to the left due to the zero point energy. Note that, the transmittivity and reflectivity curves shown in Fig. 1(c) only applies to the energy range of $E>U+1$ for every given step height $U$. It is seen that with increase in the normalized plasmon energy $E$ the reflectivity/tramsmittivity over the potential step decreases/increases. On the other hand, for the reflection/transmition of a plasmon over a negative potential step shown in Fig. 1(d) for all depth values of the potential step the transmission occurs only beyond the zero point energy value, $E=1$.

Figure 2(a) and 2(b) show the electrostatic potential and wave function profiles for the plasmon excitations at beating energy level ($E\simeq 1$) interacting with the negative potential steps, $U=-2$. It is remarked that the plasmon wavenumber increases in the potential step region while the amplitude of oscillations decreases. It also noted that the two tone character of plasmon is retained in the transmitted region. Figures 2(c) and 2(d) show the plasmon electrostatic potential and wave function profiles as interacting with the positive potential step, $U=2$. It is remarked that the wavenumber of transmitted plasmon electrostatic potential and wave function both decreases, whereas, their amplitudes of oscillation increases. This is obviously opposite to the case with the negative step potential. The later is due to the fact that the energy of plasmon depends on the wavenumber in the transmitted region through, $k=\sqrt{2m(E-U)}/\hbar$. Note that for single particle tunneling the transmission condition is $E>U$ which is different from that of the plasmon stated above.

\section{Interaction with Finite Potential Barrier/Well}

We proceed with investigation of the plasmon interaction with a rectangular potential barrier/well of normalized height/depth $U$ and width $a$. The electrostatic potential and wave function in three different regions of $x<0$, $0<x<a$ and $x>a$ are assumed to be
\begin{subequations}\label{wp2s}
\begin{align}
&{\Phi _1}(x) = \frac{{\left( {k_{21}^2{\Phi _0} + {\Psi _0}} \right)\exp \left( {{\rm{i}}{k_{11}}x} \right) - \left( {k_{11}^2{\Phi _0} + {\Psi _0}} \right)\exp \left( {{\rm{i}}{k_{21}}x} \right)}}{{2{\alpha _1}}},\\
&{\Phi _2}(x) = \frac{{\left( {k_{22}^2{\Phi _0} + {\Psi _0}} \right)\exp \left( {{\rm{i}}{k_{12}}x} \right) - \left( {k_{12}^2{\Phi _0} + {\Psi _0}} \right)\exp \left( {{\rm{i}}{k_{22}}x} \right)}}{{2{\alpha _2}}},\\
&{\Phi _3}(x) = \frac{{\left( {k_{21}^2{\Phi _a} + {\Psi _a}} \right)\exp \left( {{\rm{i}}{k_{11}}x} \right) - \left( {k_{11}^2{\Phi _a} + {\Psi _a}} \right)\exp \left( {{\rm{i}}{k_{21}}x} \right)}}{{2{\alpha _1}}}, \\
&{\Psi _1}(x) = \frac{{\left( {k_{21}^2{\Psi _0} + {\Phi _0}} \right)\exp \left( {{\rm{i}}{k_{21}}x} \right) - \left( {k_{11}^2{\Psi _0} + {\Phi _0}} \right)\exp \left( {{\rm{i}}{k_{11}}x} \right)}}{{2{\alpha _1}}}\\
&+ r\frac{{\left( {k_{21}^2{\Psi _0} + {\Phi _0}} \right)\exp \left( { - {\rm{i}}{k_{21}}x} \right) - \left( {k_{11}^2{\Psi _0} + {\Phi _0}} \right)\exp \left( { - {\rm{i}}{k_{11}}x} \right)}}{{2{\alpha _1}}},\\
&{\Psi _2}(x) = A\frac{{\left( {k_{22}^2{\Psi _0} + {\Phi _0}} \right)\exp \left( {{\rm{i}}{k_{22}}x} \right) - \left( {k_{12}^2{\Psi _0} + {\Phi _0}} \right)\exp \left( {{\rm{i}}{k_{12}}x} \right)}}{{2{\alpha _2}}}\\
&+ B\frac{{\left( {k_{22}^2{\Psi _0} + {\Phi _0}} \right)\exp \left( { - {\rm{i}}{k_{22}}x} \right) - \left( {k_{12}^2{\Psi _0} + {\Phi _0}} \right)\exp \left( { - {\rm{i}}{k_{12}}x} \right)}}{{2{\alpha _2}}},\\
&{\Psi _3}(x) = t\frac{{\left( {k_{21}^2{\Psi _0} + {\Phi _0}} \right)\exp \left( {{\rm{i}}{k_{21}}x} \right) - \left( {k_{11}^2{\Psi _0} + {\Phi _0}} \right)\exp \left( {{\rm{i}}{k_{11}}x} \right)}}{{2{\alpha _1}}},
\end{align}
\end{subequations}
with eight unknowns $A$, $B$, $r$, $t$, $\Psi_0$, $\Phi_0$, $\Psi_a$ and $\Phi_a$ to be defined from eight following boundary conditions at the barrier edges $x=0$ and $x=a$
\begin{subequations}\label{bc2}
\begin{align}
&{\Phi _1}(0) = {\Phi _2}(0),\hspace{3mm}{\left. {\frac{{d{\Phi _1}(x)}}{{dx}}} \right|_{x = 0}} = {\left. {\frac{{d{\Phi _2}(x)}}{{dx}}} \right|_{x = 0}},\\
&{\Phi _2}(a) = {\Phi _3}(a),\hspace{3mm}{\left. {\frac{{d{\Phi _2}(x)}}{{dx}}} \right|_{x = a}} = {\left. {\frac{{d{\Phi _3}(x)}}{{dx}}} \right|_{x = a}},\\
&{\Psi _1}(0) = {\Psi _2}(0),\hspace{3mm}{\left. {\frac{{d{\Psi _1}(x)}}{{dx}}} \right|_{x = 0}} = {\left. {\frac{{d{\Psi _2}(x)}}{{dx}}} \right|_{x = 0}},\\
&{\Psi _2}(a) = {\Psi _3}(a),\hspace{3mm}{\left. {\frac{{d{\Psi _2}(x)}}{{dx}}} \right|_{x = a}} = {\left. {\frac{{d{\Psi _3}(x)}}{{dx}}} \right|_{x = a}}.
\end{align}
\end{subequations}
Similar to the previous case for step potential the continuity for potential at $x=0$ always holds regardless of initial values, $\Phi_0$ and $\Psi_0$. Also, the continuity of electrostatic potential derivative at $x=0$ leads to the conclusion of $\Phi_0=\Psi_0$. However, the continuity of the electrostatic potential at $x=a$ uniquely defines the correct values of $\Phi_a$ and $\Psi_a$ as follows
\begin{subequations}\label{phiab}
\begin{align}
&{\Phi _a} = \frac{{\beta {{\rm{e}}^{ - {\rm{i}}a\left( {{k_{11}} + {k_{21}}} \right)}}\left( {1 + {k_{12}^2}} \right)\left[ {\left( {{k_{11}} - {k_{22}}} \right){{\rm{e}}^{{\rm{i}}a\left( {{k_{11}} + {k_{22}}} \right)}} - \left( {{k_{21}} - {k_{22}}} \right){{\rm{e}}^{{\rm{i}}a\left( {{k_{21}} + {k_{22}}} \right)}}} \right]}}{{{{\left( {{k_{11}} - {k_{21}}} \right)}^2}\left( {{k_{11}} + {k_{21}}} \right)}}\\
&- \frac{{\beta {{\rm{e}}^{ - {\rm{i}}a\left( {{k_{11}} + {k_{21}}} \right)}}\left( {1 + {k_{22}^2}} \right)\left[ {\left( {{k_{11}} - {k_{12}}} \right){{\rm{e}}^{{\rm{i}}a\left( {{k_{11}} + {k_{12}}} \right)}} - \left( {{k_{12}} - {k_{21}}} \right)\left( {1 + {k_{22}^2}} \right){{\rm{e}}^{{\rm{i}}a\left( {{k_{12}} + {k_{21}}} \right)}}} \right]}}{{{{\left( {{k_{11}} - {k_{21}}} \right)}^2}\left( {{k_{11}} + {k_{21}}} \right)}},\\
&{\Psi _a} = \frac{{\beta {{\rm{e}}^{ - {\rm{i}}a\left( {{k_{11}} + {k_{21}}} \right)}}\left( {1 + {k_{12}^2}} \right)\left[ {k_{11}^2\left( {{k_{21}} - {k_{22}}} \right){{\rm{e}}^{{\rm{i}}a\left( {{k_{21}} + {k_{22}}} \right)}} - k_{21}^2\left( {{k_{11}} - {k_{22}}} \right){{\rm{e}}^{{\rm{i}}a\left( {{k_{11}} + {k_{22}}} \right)}}} \right]}}{{{{\left( {{k_{11}} - {k_{21}}} \right)}^2}\left( {{k_{11}} + {k_{21}}} \right)}}\\
&+ \frac{{\beta {{\rm{e}}^{ - {\rm{i}}a\left( {{k_{11}} + {k_{21}}} \right)}}\left( {1 + {k_{22}^2}} \right)\left[ {k_{21}^2\left( {{k_{11}} - {k_{12}}} \right){{\rm{e}}^{{\rm{i}}a\left( {{k_{11}} + {k_{12}}} \right)}} + k_{11}^2\left( {{k_{12}} - {k_{21}}} \right)\left( {1 + {k_{22}}^2} \right){{\rm{e}}^{{\rm{i}}a\left( {{k_{12}} + {k_{21}}} \right)}}} \right]}}{{{{\left( {{k_{11}} - {k_{21}}} \right)}^2}\left( {{k_{11}} + {k_{21}}} \right)}}.
\end{align}
\end{subequations}
The electrostatic potential and wave function in different regions are then given as
\begin{subequations}\label{wp2n}
\begin{align}
&{\Phi _1}(x) = \frac{{\left( {1 + k_{21}^2} \right){{\rm{e}}^{{\rm{i}}{k_{11}}x}} - \left( {1 + k_{11}^2} \right){{\rm{e}}^{{\rm{i}}{k_{21}}x}}}}{{2{\alpha _1}}},\\
&{\Phi _2}(x) = \frac{{\left( {1 + k_{22}^2} \right){{\rm{e}}^{{\rm{i}}{k_{12}}x}} - \left( {1 + k_{12}^2} \right){{\rm{e}}^{{\rm{i}}{k_{22}}x}}}}{{2{\alpha _2}}},\\
&{\Phi _3}(x) = \frac{{\left( {k_{21}^2{\Phi _a} + {\Psi _a}} \right){{\rm{e}}^{{\rm{i}}{k_{11}}x}} - \left( {k_{11}^2{\Phi _a} + {\Psi _a}} \right){{\rm{e}}^{{\rm{i}}{k_{21}}x}}}}{{2{\alpha _1}}},\\
&{\Psi _1}(x) = \frac{{\left( {1 + k_{21}^2} \right){{\rm{e}}^{{\rm{i}}{k_{21}}x}} - \left( {1 + k_{11}^2} \right){{\rm{e}}^{{\rm{i}}{k_{11}}x}}}}{{2{\alpha _1}}} + r\frac{{\left( {1 + k_{21}^2} \right){{\rm{e}}^{ - {\rm{i}}{k_{21}}x}} - \left( {1 + k_{11}^2} \right){{\rm{e}}^{ - {\rm{i}}{k_{11}}x}}}}{{2{\alpha _1}}},\\
&{\Psi _2}(x) = A\frac{{\left( {1 + k_{22}^2} \right){{\rm{e}}^{{\rm{i}}{k_{22}}x}} - \left( {1 + k_{12}^2} \right){{\rm{e}}^{{\rm{i}}{k_{12}}x}}}}{{2{\alpha _2}}} + B\frac{{\left( {1 + k_{22}^2} \right){{\rm{e}}^{ - {\rm{i}}{k_{22}}x}} - \left( {1 + k_{12}^2} \right){{\rm{e}}^{ - {\rm{i}}{k_{12}}x}}}}{{2{\alpha _2}}},\\
&{\Psi _3}(x) = t\frac{{\left( {1 + k_{21}^2} \right){{\rm{e}}^{{\rm{i}}{k_{21}}x}} - \left( {1 + k_{11}^2} \right){{\rm{e}}^{{\rm{i}}{k_{11}}x}}}}{{2{\alpha _1}}}.\\
\end{align}
\end{subequations}
The four remaining unknowns can be obtained using the boundary conditions on wave function at $x=0$ and $x=a$. For instance the continuity of wave function at $x=0$ results in $A+B=1+r$. The continuity of the wave function derivative at $x=0$ leads to the following result
\begin{equation}\label{eq2}
A - B = \frac{{\left( {1 - r} \right)\left[ {{k_{11}}(1 + {k_{11}^2}) - {k_{21}}\left( {1 + {k_{21}^2}} \right)} \right]}}{{\beta \left[ {{k_{12}}(1 + {k_{12}^2}) - {k_{22}}\left( {1 + {k_{22}^2}} \right)} \right]}}.
\end{equation}
Moreover, the continuity of wavefunction and its derivative at $x=a$ lead to the following relations
\begin{subequations}\label{eq56}
\begin{align}
&t = \beta \frac{{\left( {1 + k_{22}^2} \right)\left( {A{{\rm{e}}^{{\rm{i}}a{k_{22}}}} + B{{\rm{e}}^{ - {\rm{i}}a{k_{22}}}}} \right) - \left( {1 + k_{12}^2} \right)\left( {A{{\rm{e}}^{{\rm{i}}a{k_{12}}}} + B{{\rm{e}}^{ - {\rm{i}}a{k_{12}}}}} \right)}}{{\left( {1 + k_{21}^2} \right){{\rm{e}}^{{\rm{i}}a{k_{21}}}} - \left( {1 + k_{11}^2} \right){{\rm{e}}^{{\rm{i}}a{k_{11}}}}}},\\
&t = \beta \frac{{{k_{12}}\left( {1 + {k_{12}^2}} \right)\left( {A{{\rm{e}}^{{\rm{i}}a{k_{12}}}} - B{{\rm{e}}^{ - {\rm{i}}a{k_{12}}}}} \right) - {k_{22}}\left( {1 + {k_{22}^2}} \right)\left( {A{{\rm{e}}^{{\rm{i}}a{k_{22}}}} - B{{\rm{e}}^{ - {\rm{i}}a{k_{22}}}}} \right)}}{{{k_{11}}\left( {1 + {k_{11}^2}} \right){{\rm{e}}^{{\rm{i}}a{k_{11}}}} - {k_{21}}\left( {1 + {k_{21}^2}} \right){{\rm{e}}^{{\rm{i}}a{k_{21}}}}}},
\end{align}
\end{subequations}
from which the four unknowns $A$, $B$, $r$ and $t$ can be derived, Here the long expressions for these quantities are omitted for simplicity.

\textbf{The transmittivity of a plasmon excitations from potential barrier/well, $U$, is shown for different values of the well width and barrier height, in Fig. 3. The quantum barrier/well may be either realized by two parallel metallic plates separated by a micron sized insulator gap, the so-called Josephson junction, or a heavily doped p-n-p (n-p-n) tunneling transistor junction. The quantum interference effect is clearly evident in the oscillatory transmission amplitude profiles for all given values of plasmon parameters. These profiles are exactly analogous to the case of single quantum particle tunneling through the potential barrier/well. Figure 3(a) shows the transmittivity for potential height/depth $U=-8,8$ in units of $2\epsilon_p$ and width $a=4$ in unit of the characteristic plasmon wavelength $\lambda_p=h/\sqrt{2m\epsilon_p}$ with $h$ being the Planck constant. It is observed that the transmittivity periodically reaches the maximum value of $T=1$ the value of normalized plasmon energy $E$ is increased. It is also remarked that for a barrier beyond the cutoff transmission energy much rapidly compared to that of the well with the same width. However, the amplitude of oscillations in transmittivity decreases rapidly with increase of the incident plasmon energy value. Figure 3(b) shows the transmittivity profile for the same potential width but enhanced barrier/well height/depth. It is seen that the tramsmitivity minima fall shallower and the oscillation frequency decreases significantly as the potential hiegth/depth increases. Moreover, a slight increase in the barrier/well width in Fig. 3(c) leads to sharp increase in the oscillation frequency and the relative oscillation amplitude as compared to that in Fig. 3(a). A two-tone oscillation in transmittivity profile of the free electron plasmon transmission over both potential barrier and well is clearly evident in Fig. 3(c) which is the result of the two scale length character of plasmon excitations. Finally, the increase in the potential width in Fig. 3(d) leads to sharp in oscillation frequency in the transmittivity profiles of both potential barrier and well.}

Figure 4 shows the wave function profiles for plasmon excitation with the given energy interacting with the potential well of height/depth, $U$, and width, $a$. The Fig. 4(a) shows the corresponding profiles at the beating energy level, $E\simeq 1$ with enhanced probability amplitudes in sides with no potential energy. However, as it is remarked, the beating effect is destroyed in the middle region where the well potential is present. With increase in the value of the well depth $U$ in Fig. 4(b) the wavenumber inside the potential well increases. Moreover, Fig. 4(c) shows that a slight increase in the value of normalized plasmon energy $E$ rapidly destroys the beating structures in electrostatic potential and wave function profiles leading to a slightly higher wavenumber in the potential region. Also, the increase in the width of potential well in Fig. 4(d) is shown to significantly increase the wavenumber value in the potential well region.

Figure 5, on the other hand, shows the corresponding electrostatic potential profiles with the potential well depth and energy values used in Fig. 4. It is seen that the energy value of $E\simeq 1$ caused a beating effect on the profile of electrostatic profile similar to the case of wave function profiles in Figs. 4(a) and 4(b). It is remarked that the presence of potential well leads to a much weaker oscillations in electrostatic potential in the potential well region compared to that for wave function variations shown in Fig. 4. On the other hand, comparison of Figs. 5(a) and 5(b) reveals that increase in the potential well depth leads to increase in the wavenumber of oscillations in the middle region similar to the case with wave function profile in Figs. 4(a) and 4(b). It is observed that a slight increase in the value of plasmon energy in Fig. 5(c) completely destroys the beating condition in Figs. 5(a) and 5(b). Figure 5(d) shows the effect of increase in the width of the potential well on electrostatic potential variation in the scattered regions.

Figure 6 shows the scattered wave function and electrostatic potential profiles of plasmon interacting with the potential barrier of height $U=2$ and width $a$. Figures 6(a) and 6(b) show these profiles for the plasmon energy value of $E\simeq U+1=3.01$. It is remarked that this value of the normalized plasmon energy leads to the beating condition inside the barrier opposite to the case with the potential well. On the other hand, Figs. 6(a) and 6(b) depict the scattering amplitudes for the wave function and electrostatic potential with increased plasmon energy value of $E=4$. It is remarked that the amplitude of transmissions are substantially lower compared to those of the beating energy condition in Figs. 6(a) and 6(b).

\section{Conclusion}

We studied the propagation and interaction of plasmon excitations through the positive/negative potential step and barrier/well using the Schr\"{o}dinger-Poisson model.  A coupled pseudoforce system of differential equations were deduced from which the generalized plane wave solution for plasmon wave function and electrostatic potential was derived. It was remarked that free electron plasmon excitations have two distinct characteristic length scales corresponding to single particle oscillations and collective excitations. The later aspect of plasmon excitations makes them unique and more exotic compared to free single electron case. The zero-point energy for plasmon excitation is found to be $\epsilon_0=\mu_0+2\epsilon_p$ in which $\epsilon_p$ is the plasmon energy. It is further revealed that the trasmitivity of plasmon through a potential barrier/well has oscillatory profile which is characteristic of single particle quantum tunneling through the barrier. Current investigation may be helpful for scientific developments of rapidly growing fields of plasmonics and nanotechnology.

\end{document}